\documentclass{article}
\usepackage[utf8]{inputenc}
\usepackage{amsfonts}
\usepackage{dsfont}
\usepackage[T1]{fontenc}
\usepackage{graphicx}

\usepackage{comment}

\usepackage{geometry}
\title{Representation of probability distributions with implied volatility and biological rationale}

\author{Felix Polyakov \\
\includegraphics[scale=0.65]{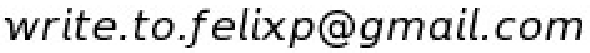} \\  \\
 Department of Mathematics\\
 Bar Ilan University\\
 Ramat-Gan, Israel}
\date{\today} 

\usepackage{graphicx}
\usepackage{amsmath}
\usepackage{cases}

\newtheorem{example}{Example}
\usepackage{graphicx}

\usepackage{eurosym}

\begin{document}

\maketitle

\begin{abstract}
\par Economic and financial theories and practice essentially deal with uncertain future. Humans encounter uncertainty in different kinds of activity, from sensory-motor control to dynamics in financial markets, what has been subject of extensive studies. Representation of uncertainty with normal or lognormal distribution is a common feature of many of those studies. For example, proposed Bayessian integration of Gaussian multisensory input in the brain or log-normal distribution of future asset price in renowned Black-Scholes-Merton (BSM) model for pricing contingent claims.
\par Standard deviation of log(future asset price) scaled by square root of time in the BSM model is called implied volatility. Actually, log(future asset price) is not normally distributed and traders account for that to avoid losses. Nevertheless the BSM formula derived under the assumption of constant volatility remains a major uniform framework for pricing options in financial markets. I propose that one of the reasons for such a high popularity of the BSM formula could be its ability to translate uncertainty measured with implied volatility into price in a way that is compatible with human intuition for measuring uncertainty.
\par The present study deals with mathematical relationship between uncertainty and the BSM implied volatility. Examples for a number of common probability distributions are presented. Overall, this work proposes that representation of various probability distributions in terms of the BSM implied volatility profile may be meaningful in both biological and financial worlds. Necessary background from financial mathematics is provided in the text.
\end{abstract}

\section{Background and introduction}
\subsection{Vanilla options}
\begin{figure}[h!]
\centering
\includegraphics[scale=0.75]{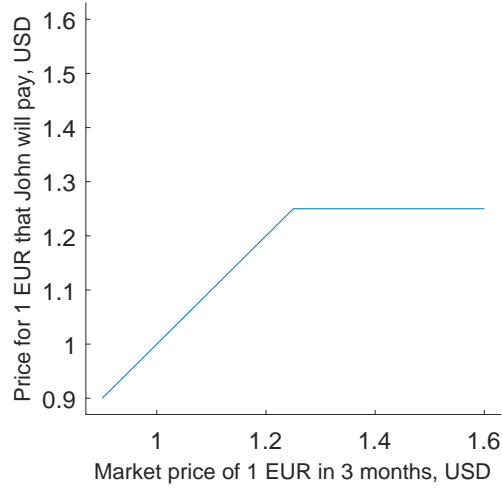}
\caption{In 3 months John will be able to buy \officialeuro100,000 for the price of \officialeuro1 as in the graph, depending on what exchange rate is in the market.}
\label{fig:call.payout.exaple}
\end{figure}

\begin{example}\label{example:call.option}
\par Today \officialeuro1 costs \$1.25. John wants to buy \officialeuro100,000 in 3 months from today. John does not want to bear a risk that the price of Euro will go up, on the other hand he wants to benefit in case the price of Euro will go down. To allow this, John buys a 3 months \emph{European call option} with strike \$1.25. If in 3 months \officialeuro1 costs \$1.20, John will pay \$120,000 for his \officialeuro100,000, less than \$125,000 that he would pay today. If in 3 months \officialeuro1 costs \$1.30, John will exercise his option and buy \officialeuro100,000 for \$125,000 instead of \$130,000. The graph with the amount John will pay for \officialeuro1 depending on future \officialeuro/\$ exchange rate is depicted in Figure \ref{fig:call.payout.exaple}. If \officialeuro/\$ exchange rate is above the strike at option's 3 months expiry John can simply benefit from the difference instead of taking a position in \officialeuro\footnote{Say, 1\officialeuro will cost 1.35\$ in 3 months. Then John will be able to benefit $(1.35\$ - 1.25\$) \times 100,000 = \$10,000$.}.
\end{example}

\par As demonstrated in Example \ref{example:call.option}, European call option provides an insurance against price increase at option's expiry above a strike price defined in option's contract. Option holder has a right to benefit from certain market moves and at the same time bears no obligations. Therefore, option is not free and has to be purchased as any insurance. If John from Example \ref{example:call.option} needs to sell an asset at certain future date he could use an insurance against price decrease in the form of European put option. Holder of European put option has a right but not obligation to sell an asset at the strike price at option's expiry. Both European call and European put are referred to as \emph{vanilla options}.
\par Vanilla option prices for many assets\footnote{Like exchange rates, equities, commodities, interest rates, etc.} are cited in the market. Vanilla option prices are commonly cited in terms of a notion called \emph{implied volatility}. The renowned Black-Scholes-Merton (BSM) formula for vanilla option price establishes a relationship between option price and corresponding value of implied volatility \cite{hull_book}:
\begin{equation}\label{eq:BSM.call}
    c(K, S, T) = e^{-q T} \cdot S_0 \cdot N(d_1) - e^{-r T} \cdot K \cdot N(d_2) \,,
\end{equation}
\begin{equation}\label{eq:BSM.put}
    p(K, S, T) = e^{-r T} \cdot K \cdot N(-d_2) - e^{-q T} \cdot S_0 \cdot N(-d_1)  \, ,
\end{equation}
with
\begin{equation}\label{eq:BSM.d1}
    d_1 = \frac{\ln(S_0 / K) + \left(r - q + \sigma^2 / 2 \right) \cdot T}{\sigma \cdot \sqrt{T}}  \, ,
\end{equation}
\begin{equation}\label{eq:BSM.d2}
    d_2 = \frac{\ln(S_0 / K) + \left(r - q - \sigma^2 / 2 \right) \cdot T}{\sigma \cdot \sqrt{T}} = d_1 - \sigma \cdot \sqrt{T}\, ,
\end{equation}
\begin{equation}\label{eq:BSM.N}
    N(x) = \frac{1}{\sqrt{2 \pi}} \int_{-\infty}^{x}   e^{-y^2/2} dy \, .
\end{equation}
The following notation is used in formulae \eqref{eq:BSM.call} - \eqref{eq:BSM.N}:
\begin{itemize}
  \item $c$ - price of vanilla call option
  \item $p$ - price of vanilla put option
  \item $S_0$ - today's asset price
  \item $K$ - strike price of the asset in option's contract
  \item $T$ - time to option's expiry, i.e. when asset's price is compared to the strike
  \item $r$, $q$ - local (of the currency that measures the asset price) and foreign interest\footnote{Intuitively, say one deposits in a bank \$1,000 and get \$1,050 in 1 year. In this case one earns 5\% interest, in other words the interest rate of the deposit is 5\%.} respectively. In case of the option from Example \ref{example:call.option} USD is local currency and EUR is foreign currency
  \item $\sigma$ - \emph{implied volatility}; when $T$ is fixed and value of $\sigma$ is the same for any strike $K$, $\sigma$ is equal to the standard deviation of annualized continuous returns\footnote{Standard deviation of $\ln\left(S_T / S\right) / \sqrt{T}$, where $S_T$ is \emph{unknown} asset's price at future time $T$.} of $S_T$
\end{itemize}
Here parameters $K$ and $T$ are defined in option's contract, $S$, $r$ and $q$ are cited in the market. Values of $r$, $q$ depend on $T$\footnote{One may deposits \$1,000 for 1 year and get 5\% yearly interest, she may deposit \$1,000 for 2 years and get a higher 6\% yearly interest (receive $1.06^2 \cdot \$1,000$) in 2 years.}. For the same $K$, $T$ both call $c$ and put $p$ will imply the same value of implied volatility $\sigma$; otherwise situation will create an arbitrage opportunity in the market and arbitrage will be realized pretty fast.

\par There is one-to-one (and of course monotonous) correspondence between vanilla option prices ($c$, $p$) and values of implied volatility ($\sigma$). Consider an ideal and non-realistic case: for some predefined $T$ and for an arbitrary $K$ the same values of implied volatility $\sigma$ are implied from option prices cited in the market, then the uncertainty of $\ln S_T$ is described by normal distribution with standard deviation $s = \sigma \sqrt{T}$ and expectation $\mu = \ln S + (r - q - \sigma^2/2) T$. Equivalently, the distribution of the future asset price is lognormal with the probability density function
\begin{equation}\label{eq:lognormal.distribution}
    p(x) = \frac{1}{x s \sqrt{2 \pi}}e^{-\frac{(\ln x - \mu)^2}{2 s^2}}\, .
\end{equation}
The book by J. Hull "Options, futures and other derivatives" is among many different references with more detailed exposition of the basics of financial mathematics.

\subsection{Vanilla price and implied probability density function}

\par Let $K$, $T$, $S$ be option's strike, time to expiry and today's spot respectively. Let us write the formula for a call price in the form
\begin{equation}\label{eq:callprice}
    c(K, S, T) = e^{-r T} \int_{K}^{\infty}\left(S_{T} - K \right) \cdot p(S_T) d S_T \, ,
\end{equation}
where $p(S_{T})$ denotes implied\footnote{The word ``implied'' will sometimes be omitted further in text. Computations of volatility smiles in this work are related exclusively to implied probabilities.} probability density of asset price at option's expiry and $P(S_T)$ denotes corresponding cumulative probability.
Consequently,
\begin{equation}\label{eq:1st.derivative.call}
    \displaystyle\frac{\partial c(K, S, T)}{\partial K} = -e^{-r T} \int_{K}^{\infty}p(S_T) d S_T = -e^{-r T} \left[1 - P(K)\right] = e^{-r T} \left[ P(K) - 1 \right]\, .
\end{equation}
Hence
\begin{equation}\label{eq:2nd.derivative.call}
    \displaystyle\frac{\partial^2 C(K, S, T)}{\partial K^2} = e^{-r T} p(K)\, .
\end{equation}
\par In the ideal case when distribution of the future asset price $p(S_T)$ is lognormal, formula \eqref{eq:callprice} becomes \cite{hull_book}:
\begin{eqnarray}
\nonumber
    c(X, S, T) & = & \frac{e^{-r T}}{\sigma \sqrt{2 \pi}} \int_{K}^{\infty}\frac{S_{T} - K }{S_T} \exp\left(-\frac{\ln S_T - \mu }{2 \sigma^2}\right) d S_T = \\
    \nonumber
    & = & \int_{(\ln K - \mu) / \sigma}^{\infty} \left( e^{Q \sigma + \mu} - K \right) \cdot \left(\frac{1}{\sqrt{2 \pi}} e^{-Q^2 / 2} \right) dQ\, ,
\end{eqnarray}
with $Q = \frac{\ln S_T - \mu}{\sigma}$; $\mu$, $\sigma$ being respectively the mean and standard deviation of $\ln S_T$.
For lognormal $p(S_T)$ formula \eqref{eq:callprice} is equivalent to formula \eqref{eq:BSM.call} \cite{hull_book}.
\par The formula for the put option price is:
\begin{equation}\label{eq:putprice}
    p(K, S, T) = e^{-r T} \int_{-\infty}^{K}\left(K - S_{T}\right) \cdot p(S_T) d S_T \, .
\end{equation}

The put-call parity states the following relationship between the prices of put and call with the same strike:
\begin{equation}\label{eq:put_call_parity}
      e^{r T} \left(\mbox{call}(K, S, T) - \mbox{put}(K, S, T) \right) = \int_{-\infty}^{\infty} \left(S_{T} - K \right)\cdot p(S_T) d S_T = \mathbb{E}(S_T) - K = \mbox{ATMF} - K \, .
\end{equation}

\subsection{Biological motivation for representing uncertainty with log-normal implied volatility}
\par According to Weber-Fechner law, the subjective perception/sensation is proportional to the logarithm of the stimulus intensity \cite{Fechner:1860}:
\begin{equation}\label{eq:weber-fechner.law}
    p = \kappa \ln \frac{S}{S_0}\, ,
\end{equation}
$p$ is perception, $S$ is stimulus intensity and $\kappa$ is constant.
In other words, the relationship between stimulus and perception is logarithmic. Correspondingly, it is reasonable to propose that normal uncertainty in the stimulus would result in log-normal uncertainty in perception of the stimulus.
It has been shown that the population distribution of the intrinsic excitability or gain of a neuron is a heavy tail distribution, more precisely a log-normal shape \cite{Scheler:2017}.
\par Different empirical works provided evidence that the brain both represents probability distributions and performs probabilistic inference, see for example review \cite{Pouget.Beck.Ma.Latham:2013}. A study of uncertainty learning and integration in sensorimotor learning showed that subjects internally represent statistical distribution of task's uncertainty ($p_{prior}$) and subject's uncertainty about true value of the sensed input ($p(x_{true}|x_{sensed})$); the authors concluded that the central nervous system employs probabilistic models during sensorimotor learning \cite{kording.wolpert:2004}. The study employed bayesian integration of normally distributed uncertainties. Based on Weber-Fechner law \cite{Fechner:1860}, it could be proposed that normally distributed uncertainty of what is sensed/produced leads to log-normal type of uncertainty about the outer world within the internal representation.
\par For different kinds of assets, for example FX exchange rates, traders mostly measure option prices in terms of log-normal implied volatility instead of price measured in terms of currency. That is traders and financial institutions naturally prefer to use implied volatility as alternative measure for pricing uncertainty of the future asset price. Market's estimates for uncertainty of future asset prices deviate from log-normal as vanilla options with different strikes $K$ have different implied volatilities. Nevertheless, the uncertainty is measured in a sense as a deviation from log-normal distribution whose implied volatility is constant. The relationship $\sigma(K)$ between implied volatility and option's strike is called \emph{volatility smile}.

\section{Results}
\par Any probability distribution defined by the density function $p(x)$ whose integral \eqref{eq:callprice} exists for any $K$ can be represented with implied volatility smile. Below a process for obtaining representation of probability distribution with implied volatility smile is demonstrated for a number of commonly used distributions with graphical examples of volatility smiles. In general the algorithm is as follows:
\begin{enumerate}
    \item Given $p(S_T)$ use \eqref{eq:callprice} to compute vanilla option prices for a range of strikes.
    \item Use option prices to compute implied volatility for selected strikes. Volatilities are implied from formula \eqref{eq:BSM.call} by numerical procedure for solving non-linear equations. Among the simplest is half-division rule.
\end{enumerate}

\subsection{Implied volatility for different probability distributions}
\par
\begin{figure}[h!]
\centering 
\includegraphics[scale=0.8]{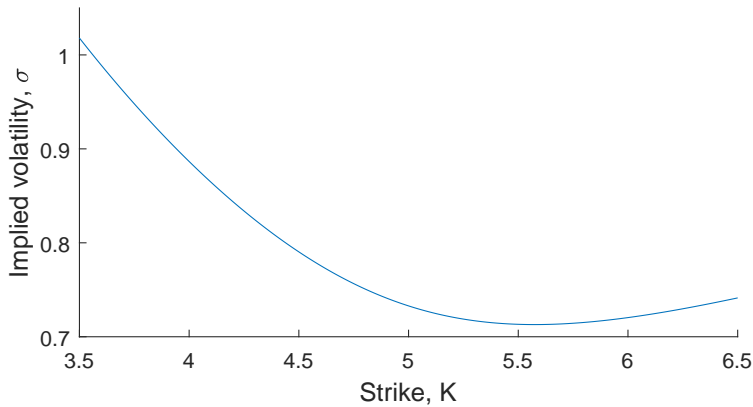}
\caption{Options with different strikes ranging from \$3.5 to \$6.5 per unit of asset expire in 6 months. Asset price in 6 month is unknown but it is known that implied probability distribution is Student's $t$ with $\nu = 1.5$ and its expected value is translated from zero to \$5. For such a setting the implied volatility profile is depicted in the figure; the profile is not flat. Derivations and formulae related to option prices on an asset whose future price follows translated Student's t-distributions are provided further in text.}
\label{fig:vol.smile.exaple}
\end{figure}

 An example of implied volatility profile as a function of strike $K$ is demonstrated in Figure \ref{fig:vol.smile.exaple}. The implied volatility in the figure is not flat and so the risk-neutral probability density for the future asset price is not log-normal. The graph of  volatility smile shows that the distribution has fatter tails\footnote{The higher is implied volatility the greater is uncertainty and consequently vanilla option is more expensive.} on both sides (especially on the left side) compared to log-normal distribution with $\sigma$ equal to minimal (on the graph) value of implied volatility.

\par Probability distribution is defined by the shape of the volatility smile $\sigma(K)$, spot $S$, time to maturity $T$, interest rates $r$ and $q$. Values of $S$, $r$, $q$, $T$ can be used to compute expectation of the future asset price equal to ATMF and to put a constraint on parameters of probability density function. Below I use vanilla call prices to compute implied volatility profiles that correspond to a number of implied probability density functions listed in Table \ref{tab:formulae.expectation.callprice}.

\begin{table}
\caption{Formulae for the expected value and call price for different risk-neutral probability distributions.}
\begin{tabular}{|c|c|c|c|}
  \hline \hline
  Distribution  & Density function & Expected value (= ATMF = $S_0 \cdot e^{(r - q) T}$)   & Call price \\
  \hline  \hline
  Lognormal          &   $\frac{1}{x s \sqrt{2 \pi}}e^{-\frac{(\ln x - \mu)^2}{2 s^2}}$ &  $e^{\mu + \frac{s^2}{2}},\; \mu = \ln S_0 + (r - q) T - \frac{s^2}{2}$                     & formula \eqref{eq:BSM.call} \\
  Gamma              & $\frac{x^{\kappa-1} e^{-\frac{x}{\theta}}}{\theta^{\kappa} \Gamma(\kappa)}$ & $\kappa \cdot \theta$    & formula \eqref{eq:callprice.gamma} \\
  Normal             &   $\frac{1}{s \sqrt{2 \pi}}e^{-\frac{( x - \mu)^2}{2 s^2}}$ &    $\mu$                    & formula \eqref{eq:normal.call.price} \\
  Translated Student & $\frac{\Gamma\left( \frac{\nu + 1}{2} \right)}{\sqrt{\nu \pi}\, \Gamma \left( \frac{\nu}{2} \right)} \left( 1 + \frac{\left(x - \mu \right)^2}{\nu} \right) ^ {-\frac{\nu + 1}{2}}$ & $\mu$                    & formula \eqref{eq:call.student.final} \\
  Uniform            & $\begin{cases}
   \frac{1}{b - a} &  a \leq x \leq  b \\
   0       & \mbox{otherwise}
  \end{cases}$ &  $\frac{a + b}{2}$        & formula \eqref{eq:unifordistr.callprice} \\
  Log uniform        & $\begin{cases}
   \frac{1}{x (\ln b - \ln a)} &  a \leq x \leq  b \\
   0       & \mbox{otherwise}
  \end{cases}$ & $\frac{b - a}{\ln b - \ln a}$     & formula \eqref{eq:logunifordistr.callprice} \\
  \hline
\end{tabular}
\label{tab:formulae.expectation.callprice}
\end{table}

\subsection{Gamma distribution of asset price at expiry}
\par The probability density function of gamma distribution depends on two parameters, $\kappa$ and $\theta$:
\begin{equation}\label{eq:gamma.distribution}
    p(x;\, \kappa,\, \theta) = \frac{x^{\kappa-1} e^{-\frac{x}{\theta}}}{\theta^{\kappa} \Gamma(\kappa)},\;\; x > 0,\; \kappa,\, \theta > 0\, .
\end{equation}
Now compute the price of a call option when the risk neutral probability density follows gamma distribution.
The computation will use the definition \eqref{eq:callprice} of a call price:
\begin{eqnarray}
    \nonumber
    \mbox{call}(K, r, T) & = & e^{-r T} \int_{K}^{\infty}\left(S_{T} - K \right) \cdot p(S_T;\, \kappa,\, \theta) d S_T  = \\
    \nonumber
    & & e^{-r T} \int_{X}^{\infty}\left\{ \frac{{S_{T}}^{\kappa} e^{-\frac{S_{T}}{\theta}}}{\theta^{\kappa} \Gamma(\kappa)} - K  p(S_{T};\, \kappa,\, \theta) \right\} d S_{T} = \\
    \nonumber
    & & e^{-r T} \int_{X}^{\infty}\left\{ \theta \cdot \kappa \cdot \frac{{S_{T}}^{\kappa} e^{-\frac{S_{T}}{\theta}}}{\theta^{\kappa + 1} \Gamma(\kappa + 1)} - K  p(S_{T};\, \kappa,\, \theta) \right\} d S_{T} = \\
     \label{eq:callprice.gamma}
     & & e^{-r T} \left\{\theta \cdot \kappa \cdot \left[1 - P(K; \kappa + 1, \theta) \right] - K \cdot \left[1 - P(K; \kappa, \theta) \right] \right\}
    \, .
\end{eqnarray}
Here $P(K; \kappa, \theta)$ is the cumulative probability density function of gamma distribution \eqref{eq:gamma.distribution}.
Correspondingly, the at the money forward (ATMF) of the asset would be the expected value of risk neutral probability density and for gamma distribution the ATMF is:

\begin{eqnarray}
   \nonumber
   ATMF & = & \int_{0}^{\infty} S_{T} \cdot p(S_T;\, \kappa,\, \theta) d S_T = \\
   & &
   \label{eq:ATMF.gamma}
   \int_{0}^{\infty}\frac{{S_{T}}^{\kappa} e^{-\frac{S_{T}}{\theta}}}{\theta^{\kappa} \Gamma(\kappa)} d S_{T} = \theta \Gamma(\kappa+1) / \Gamma(\kappa) = \kappa \theta \, .
\end{eqnarray}
By using the formula for the call price \eqref{eq:callprice.gamma} and the put-call parity \eqref{eq:put_call_parity} get the price of the put option:
\begin{equation}\label{eq:ATMF.based.on.spot}
   \mbox{put}(K,\, T) = \mbox{call}(K,\, T) - e^{-r T} \left(\mbox{ATMF} - K\right) = e^{-r T} \left[\theta \kappa \cdot P(K; \kappa + 1, \theta)  + K \cdot P(K; \kappa, \theta)  \right] \; .
\end{equation}

\par Formula \eqref{eq:ATMF.gamma} will be useful for constraining parameters of the risk-neutral probability density function $p$ given that
\begin{equation}\label{eq:ATMF.based.on.spot}
   ATMF \equiv S_{0} \cdot e^{\left(r_{T} - q_T\right) T}\, .
\end{equation}
So the spot value now satisfies the following equality:
\begin{equation}\label{eq:spot.based.on.ATMF.gamma}
   S_0 = ATMF \cdot e^{-\left(r_{T} - q_T\right) T} = \kappa \theta  \cdot e^{-\left(r_{T} - q_T\right) T}\, .
\end{equation}
Equivalently,
$\kappa \theta = S_{0}\cdot e^{\left(r_{T} - q_T\right) T} $ or
\begin{eqnarray}
  \label{eq:kappa.of_gamma.functionof.spot}
  \kappa &=& S_{0}\cdot e^{\left(r_{T} - q_T\right) T}  / \theta \, , \\
  \label{eq:theta.of_gamma.functionof.spot}
  \theta &=& S_{0}\cdot e^{\left(r_{T} - q_T\right) T}  / \kappa \, .
\end{eqnarray}
\par So, for gamma distribution
\begin{eqnarray}
    \label{eq:meangamma.of.kappatheta}
    \mbox{mean} & = & \mbox{ATMF} = \kappa \theta \, ,\\
    \label{eq:variancegamma.of.kappatheta}
    \mbox{variance} & = & \kappa \theta^2\, , \\
\label{eq:thetagamma.of.variance.mean}
   \theta & = & \frac{\mbox{variance}}{\mbox{mean}} = \frac{\mbox{variance}}{S_0 e^{(r_T - q_T) T}}\, , \\
   \label{eq:kappa.of.variancemean}
   \kappa & = & \frac{\mbox{mean}^2}{\mbox{variance}} = \frac{{S_0}^2 e^{2(r_T - q_T) T}}{\mbox{variance} } \, .
\end{eqnarray}
\par Formula \eqref{eq:theta.of_gamma.functionof.spot} can be used to introduce the value of spot $S_0$ into the formula \eqref{eq:callprice.gamma} for a call price:
\begin{flalign}\label{eq:BScall.ATMF.insteadof.kappatheta}
   & c(K, r, T)  =   e^{-r_T T} \left\{ S_{0}\cdot e^{\left(r_{T} - q_T\right) T}  \cdot \left[1 - P(K; \kappa + 1, \theta) \right] - K \cdot \left[1 - P(K; \kappa, \theta) \right] \right\} = & \\
\label{eq:BScall.S0.insteadof.theta}
    & e^{-r_T T} \left\{ S_{0}\cdot e^{\left(r_{T} - q_T\right) T}  \cdot \left[1 - P(K; \kappa + 1, S_{0}\cdot e^{\left(r_{T} - q_T\right) T}  / \kappa) \right] - K \cdot \left[1 - P(K; \kappa, S_{0}\cdot e^{\left(r_{T} - q_T\right) T}  / \kappa) \right] \right\}\, .&
\end{flalign}

\subsection{Normal distribution}
\par The case of normal implied probability of the asset price in future time $T$ corresponds to Bachelier's model of arithmetic Brownian motion \cite{bachelier:1900}.
The price of the call option according to Bachelier's formula is computed as follows \cite{dawson.blake.cairns.dowd:2007}:
\begin{eqnarray}
    \label{eq:normal.call.price}
    \mbox{call}(K, r, T) & = & e^{-r T} \left[\left(\mbox{ATMF}(T) - K \right) \cdot N\left(\frac{\mbox{ATMF}(T) - K}{\sigma_N \sqrt{T}} \right) + \sigma_N \cdot \sqrt{T} \cdot  N'\left(\frac{\mbox{ATMF}(T) - K}{\sigma_N \sqrt{T}} \right) \right] \, , \\
    \label{eq:normal.put.price}
    \mbox{put}(K, r, T) & = & e^{-r T} \left[\left(K - \mbox{ATMF}(T) \right) \cdot N\left(\frac{K - \mbox{ATMF}(T)}{\sigma_N \sqrt{T}} \right) + \sigma_N \cdot \sqrt{T} \cdot  N'\left(\frac{K - \mbox{ATMF}(T)}{\sigma_N \sqrt{T}} \right) \right] \, .
\end{eqnarray}
where $\sigma_N$ is implied \emph{normal} volatility\footnote{Equal to the standard deviation of the future asset price normalized by the square root of time in case the future asset price is normally distributed}, $F_T$ is the forward value of the asset (strike with this value nullifies the forward contract), and $N'$ is probability density of the standard normal distribution.
\par
\begin{figure}[h!]
\centering
\includegraphics[scale=0.75]{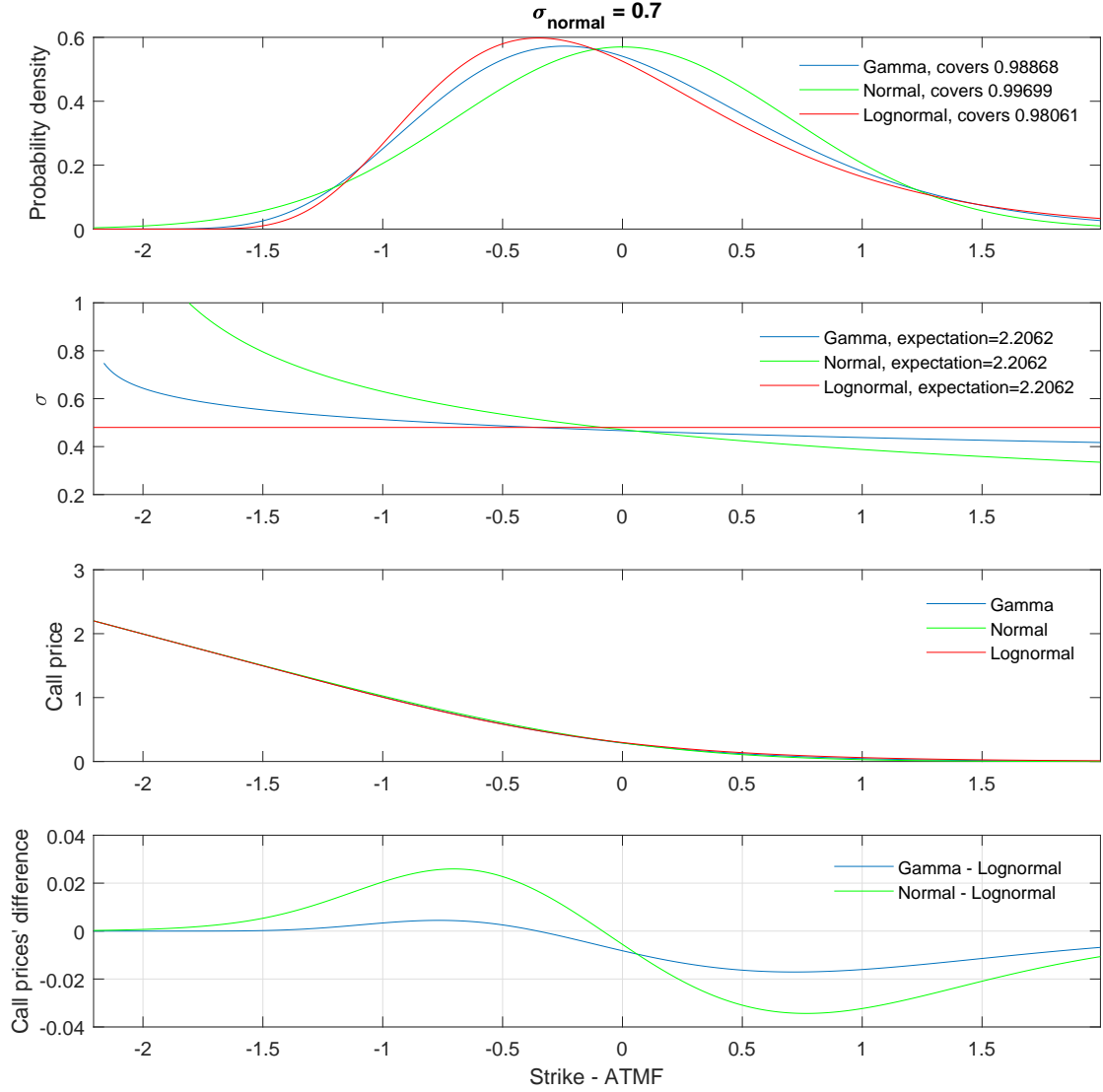} 
\caption{Three assets are considered. For one of them risk-neutral probability of future asset price is gamma, for another normal and for the third lognormal. All three distributions have the same expectation, standard deviations of lognormal and gamma distributions are adjusted to allow best possible least squares fit of their probability density functions to the normal one. Options on those 3 assets with different strikes expiry in 6 months. (Upper) Probability density functions versus Strike - expected value $\mu$. (Middle) Volatility smile. (Below) Vanilla call prices.} 
\label{fig:pdf_sigma_callprice.gamma.normal.lognormal}
\end{figure}

 Examples of probability density, volatility smile and call option prices for gamma, normal and lognormal risk-neutral probability densities are demonstrated in Figure \ref{fig:pdf_sigma_callprice.gamma.normal.lognormal}.

%

\subsection{Translated Student's $t$-distribution}
\par Probability density function for translated Student's $t$-distribution is defined as follows:
\begin{equation}\label{eq:pdf.translatedstudent.distribution}
  p(x;\, \mu,\, \nu) = \frac{\Gamma\left( \frac{\nu + 1}{2} \right)}{\sqrt{\nu \pi}\, \Gamma \left( \frac{\nu}{2} \right)} \left( 1 + \frac{\left(x - \mu \right)^2}{\nu} \right) ^ {-\frac{\nu + 1}{2}} \, .
\end{equation}
We consider the case when $\nu > 1$.
The translated distribution is obtained from Student's $t$-distribution by translating the expected value of the distribution from zero to $\mu$.

It is known that for $x \geq \mu$ the cumulative probability is as follows:
\begin{equation}\label{eq:cdf.translatedstudent.distribution.xabovemu}
  P(x;\, \mu,\, \nu) = \int_{-\infty}^{x} p(u;\, \mu,\, \nu) du = 1 - \frac{1}{2} I_{y(x)} \left( \frac{\nu}{2},\, \frac{1}{2} \right)\, ,
\end{equation}
and for $x < \mu$, using symmetry with respect to $\mu$,
\begin{equation}\label{eq:cdf.translatedstudent.distribution.xbelowmu}
  P(x;\, \mu,\, \nu) = 1 - P(2 \mu - x;\, \mu,\, \nu) = \frac{1}{2} I_{y(2 \mu - x)} \left( \frac{\nu}{2},\, \frac{1}{2} \right)\, ,
\end{equation}

where
\begin{equation}\label{y_of_x}
  y(x) = \frac{\nu}{\left(x - \mu \right)^2 + \nu}
\end{equation}
and $I$ is regularized incomplete beta function:
\begin{equation}\label{eq:incomplete.beta.function}
  I_x(a,\, b) = \frac{B(x;\, a,\, b)}{B(a,\, b)}\, ,
\end{equation}
$B(x,\, y)$ is the beta function
\begin{equation*}
  B(x,\, y) = \int_{0}^1 t^{x - 1} (1 - t)^{y - 1} dt
\end{equation*}
and
\begin{equation*}
  B(x;\, a,\, b) = \int_{0}^x t^{a - 1} (1 - t)^{b - 1} dt
\end{equation*}
is the incomplete beta function.
\par Now compute call option price for translated Student's $t$-distribution using formula \eqref{eq:callprice} and assuming $K \geq \mu$.
\begin{eqnarray}
   \nonumber
   c(K,\, \mu,\, T) & = & e^{-r T} \left[\int_{K}^{\infty} \left(S_T - \mu\right) \cdot  p(S_T;\, \mu,\, \nu) d S_T - \left(K - \mu \right) \int_{K}^{\infty} p(S_T;\, \mu,\, \nu) d S_T  \right] =
\end{eqnarray}
\begin{eqnarray}
   \nonumber
   & e^{-r T} \left\{ \frac{\Gamma\left( \frac{\nu + 1}{2} \right)}{\sqrt{\nu \pi}\, \Gamma \left( \frac{\nu}{2} \right)} \int_{K}^{\infty} \left(S_T - \mu \right) \cdot  \left( 1 + \frac{\left(S_T - \mu \right)^2}{\nu} \right) ^ {-\frac{\nu + 1}{2}} d S_T - \left(K - \mu\right) \cdot \left[1 - P(K;\, \mu,\, \nu) \right]  \right\} = \\
   \nonumber
  & e^{-r T} \left\{\frac{\nu}{2} \cdot \frac{\Gamma\left( \frac{\nu + 1}{2} \right)}{\sqrt{\nu \pi}\, \Gamma \left( \frac{\nu}{2} \right)}  \int_{K}^{\infty}\left( 1 + \frac{\left(S_T - \mu \right)^2}{\nu} \right) ^ {-\frac{\nu + 1}{2}} d \left(1 + \frac{ (S_T - \mu) ^2}{\nu} \right) - \frac{K - \mu}{2} \cdot I_{y(K)} \left( \frac{\nu}{2},\, \frac{1}{2} \right)  \right\} = \\
  \nonumber
  & e^{-r T} \left\{ \frac{\nu}{2} \cdot \frac{\Gamma\left( \frac{\nu + 1}{2} \right)}{\sqrt{\nu \pi}\, \Gamma \left( \frac{\nu}{2} \right)} \int\limits_{1 + \frac{(K - \mu)^2}{\nu}}^{\infty} w ^ {-\frac{\nu + 1}{2}} d w - \frac{K - \mu}{2} \cdot  I_{y(K)} \left( \frac{\nu}{2},\, \frac{1}{2} \right)  \right\} = \\
  \nonumber
  & e^{-r T} \left\{- \frac{\nu}{2} \cdot \frac{2}{1 - \nu} \cdot \frac{\Gamma\left( \frac{\nu + 1}{2} \right)}{\sqrt{\nu \pi}\, \Gamma \left( \frac{\nu}{2} \right)} \cdot \left[ 1 + \frac{(K - \mu)^2}{\nu} \right]^{\frac{1 - \nu}{2}} - \frac{K - \mu}{2} \cdot I_{y(K)} \left( \frac{\nu}{2},\, \frac{1}{2} \right)  \right\} = \\
  \label{eq:call.translated.student.KaboveS}
  & e^{-r T} \left\{ \frac{\nu}{\nu - 1} \cdot \frac{\Gamma\left( \frac{\nu + 1}{2} \right)}{\sqrt{\nu \pi}\, \Gamma \left( \frac{\nu}{2} \right)} \cdot \left[ 1 + \frac{(K - \mu)^2}{\nu} \right]^{\frac{1 - \nu}{2}} - \frac{K - \mu}{2} \cdot I_{y(K)} \left( \frac{\nu}{2},\, \frac{1}{2} \right)  \right\} \, .
\end{eqnarray}
Denote
\begin{eqnarray}
  \label{eq:q1.call.student}
   {q_1}^{+}(K,\, \mu) & = & e^{-r T} \frac{\nu}{\nu - 1} \cdot \frac{\Gamma\left( \frac{\nu + 1}{2} \right)}{\sqrt{\nu \pi}\, \Gamma \left( \frac{\nu}{2} \right)} \cdot \left[ 1 + \frac{(K - \mu)^2}{\nu} \right]^{\frac{1 - \nu}{2}} \, ,\\
  \label{eq:q2plus.callprice.student}
   {q_2}^{+}(K,\, \mu,\, T) & = & e^{-r T} \frac{K - \mu}{2} \cdot I_{y(K)} \left( \frac{\nu}{2},\, \frac{1}{2} \right)
\end{eqnarray}

Sign ``+'' in notation ${q_1}^{+}$ in formula \eqref{eq:q1.call.student} is used to emphasize that $K \geq \mu $.
\par Now use derivations that lead to formula \eqref{eq:call.translated.student.KaboveS} in order to compute the call price for the case $K < \mu$:
\begin{align}
  \nonumber
    {q_1}^{-}(K,\, \mu,\, T) & = e^{-r T}\left[ \int_{K}^{\mu} \left(S_T - \mu\right) \cdot  p(S_T;\, \mu,\, \nu) d S_T + \int_{\mu}^{\infty} \left(S_T - \mu\right) \cdot  p(S_T;\, \mu,\, \nu) d S_T\right] = \\
  \nonumber
   & e^{-r T} \frac{\nu}{2} \cdot \frac{\Gamma\left( \frac{\nu + 1}{2} \right)}{\sqrt{\nu \pi}\, \Gamma \left( \frac{\nu}{2} \right)} \int\limits_{1 + \frac{(K - \mu)^2}{\nu}}^{1} w ^ {-\frac{\nu + 1}{2}} d w + {q_1}^{+}(\mu,\, \mu) =  \\
   \nonumber
   & e^{-r T} \frac{\nu}{\nu - 1} \cdot \frac{\Gamma\left( \frac{\nu + 1}{2} \right)}{\sqrt{\nu \pi}\, \Gamma \left( \frac{\nu}{2} \right)} \cdot \left[\left(1 + \frac{(K - \mu)^2}{\nu} \right)^{\frac{1 - \nu}{2}} - 1 \right] + e^{-r T} \frac{\nu}{1 - \nu} \cdot \frac{\Gamma\left( \frac{\nu + 1}{2} \right)}{\sqrt{\nu \pi}\, \Gamma \left( \frac{\nu}{2} \right)} =  \\
   \label{eq:qminus.callprice.student}
  &  e^{-r T} \frac{\nu}{\nu - 1} \cdot \frac{\Gamma\left( \frac{\nu + 1}{2} \right)}{\sqrt{\nu \pi}\, \Gamma \left( \frac{\nu}{2} \right)} \cdot \left[1 + \frac{(K - \mu)^2}{\nu} \right]^{\frac{1 - \nu}{2}}\; .
\end{align}
\begin{multline}\label{eq:q2minus.callprice.student}
  {q_2}^{-}(K,\, \mu,\, T) = e^{-r T} \cdot \left(K - \mu \right) \int_{K}^{\infty} p(S_T;\, \mu,\, \nu) d S_T =
       e^{-r T} \cdot \left(K - \mu \right) \cdot \left[1 - \frac{1}{2} I_{y(2 \mu - K)} \left( \frac{\nu}{2},\, \frac{1}{2} \right) \right]\; .
\end{multline}
Due to symmetry, ${q_1}^{+} = {q_1}^{-}$.
\par Now use \eqref{eq:call.translated.student.KaboveS} - \eqref{eq:q2minus.callprice.student} to write the formula for the call price when risk-neutral probability density function is translated Student's $t$-distribution:
\begin{multline*}
    c(K,\, \mu,\, T)  =
    \begin{cases}
     {q_1}^{+}(K,\, \mu,\, T)  + {q_2}^{+}(K,\mu,T) & K  \geq  \mu \\
     {q_1}^{-}(K,\, \mu,\, T) + {q_2}^{-}(\mu,\, \mu,\, T) &  K < \mu
  \end{cases} = \\
       e^{-r T}  \frac{\nu}{\nu - 1} \cdot \frac{\Gamma\left( \frac{\nu + 1}{2} \right)}{\sqrt{\nu \pi}\, \Gamma \left( \frac{\nu}{2} \right)} \cdot \left[ 1 + \frac{(K - \mu)^2}{\nu} \right]^{\frac{1 - \nu}{2}} - e^{-r T} \frac{K - \mu}{2} \cdot
     \begin{cases}
       I_{y(K)} \left( \frac{\nu}{2},\, \frac{1}{2}\right)    & K  \geq  \mu \\
        & \\
     2 - I_{y(2\mu - K)} \left( \frac{\nu}{2},\, \frac{1}{2}\right)  &  K < \mu \; .
  \end{cases}
\end{multline*}
Noting from \eqref{y_of_x} that $y(K) = y(2 \mu - K)$ we finally get the pricing formula for the call option:
\begin{multline}\label{eq:call.student.final}
  c(K,\, \mu,\, T) =     e^{-r T}  \frac{\nu}{\nu - 1} \cdot \frac{\Gamma\left( \frac{\nu + 1}{2} \right)}{\sqrt{\nu \pi}\, \Gamma \left( \frac{\nu}{2} \right)} \cdot \left[ 1 + \frac{(\mu - K)^2}{\nu} \right]^{\frac{1 - \nu}{2}} + e^{-r T} \frac{\mu - K}{2} \cdot
     \begin{cases}
        I_{y(K)} \left( \frac{\nu}{2},\, \frac{1}{2}\right)    & K  \geq  \mu \\
        & \\
     2 -I_{y(K)} \left( \frac{\nu}{2},\, \frac{1}{2}\right)  &  K < \mu \; .
  \end{cases}
\end{multline}
Using put-call parity \eqref{eq:put_call_parity}, the price for the put option is:
\begin{multline}\label{eq:put.student.final}
    \mbox{put}(K,\, \mu,\, T) = \mbox{call}(K,\, \mu,\, T) - e^{-r T}(\mu - K)  = \\
          e^{-r T}  \frac{\nu}{\nu - 1} \cdot \frac{\Gamma\left( \frac{\nu + 1}{2} \right)}{\sqrt{\nu \pi}\, \Gamma \left( \frac{\nu}{2} \right)} \cdot \left[ 1 + \frac{(\mu - K)^2}{\nu} \right]^{\frac{1 - \nu}{2}} + e^{-r T} \frac{\mu - K}{2} \cdot
     \begin{cases}
        I_{y(K)} \left( \frac{\nu}{2},\, \frac{1}{2}\right) - 2   & K  \geq  \mu \\
        & \\
     -I_{y(K)} \left( \frac{\nu}{2},\, \frac{1}{2}\right)  &  K < \mu \; .
  \end{cases}
\end{multline}

\par
\begin{figure}[h!]
\centering
\includegraphics[scale=0.75]{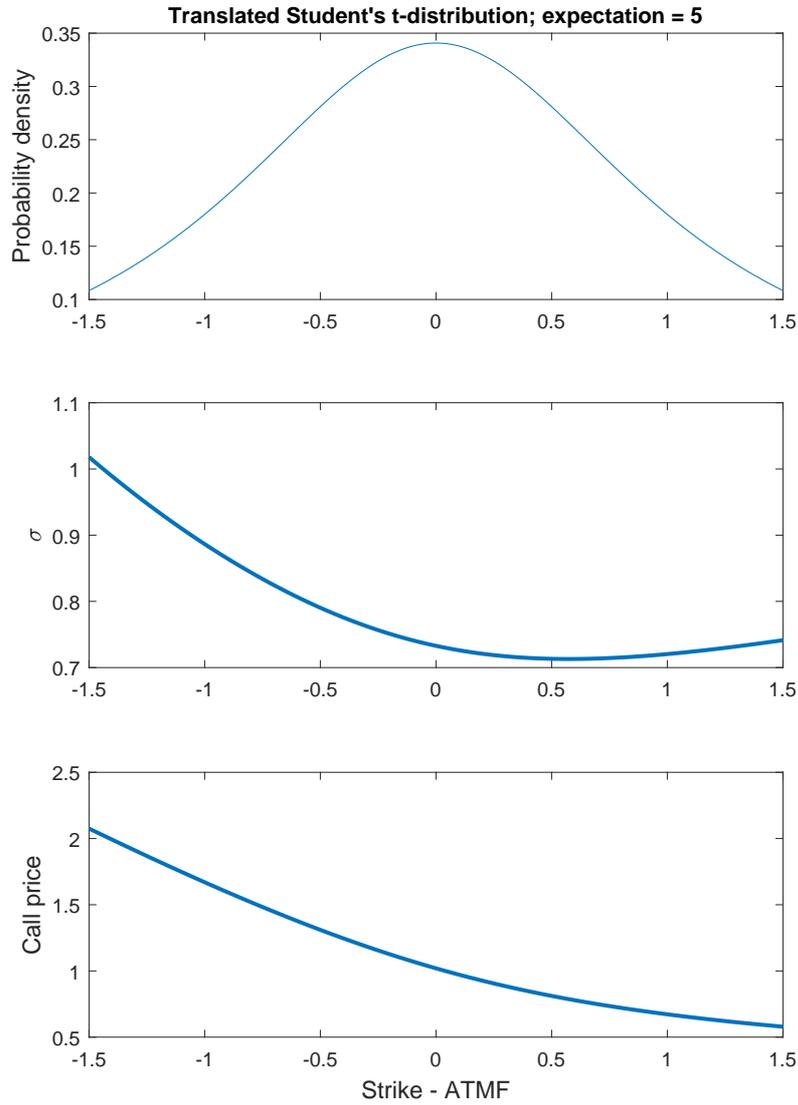}
\caption{Risk-neutral probability of future asset price is translated Student t-distribution \eqref{eq:pdf.translatedstudent.distribution}. Options with different strikes expire in 6 months. (Upper) Probability density function versus Strike - expected value $\mu$. (Middle) Volatility smile. (Below) Vanilla call prices.}
\label{fig:pdf_sigma_callprice.student}
\end{figure}

 An example of volatility smile and call prices corresponding to some Student's t-distribution is demonstrated in Figure \ref{fig:pdf_sigma_callprice.student}.

\subsection{Uniform distribution of the asset price at expiry}
\par Probability density function for uniform distribution on an interval $[a,\, b]$, $a < b$, is defined as follows:
\begin{equation}\label{eq:uniform.distribution}
 p(X;\, a,\, b) =
      \begin{cases}
   \frac{1}{b - a} & a \leq X \leq b \\
   0       & \mbox{otherwise} \ .
  \end{cases}
\end{equation}
Using equation \eqref{eq:callprice}, the corresponding call price can be written as:
\begin{eqnarray}\nonumber
    c(K;\, a,\, b) =  \frac{e^{-r T}}{b - a} \cdot
    \begin{cases}
    \int_{a}^{b}\left(S_{T} - K \right)  d S_T & K  \leq a \\
     \int_{K}^{b}\left(S_{T} - K \right)  d S_T & a < K < b \\
    0 & b \leq K
  \end{cases} = \\
  \nonumber
     \frac{e^{-r T}}{b - a} \cdot \begin{cases}
    (b^2 - a^2) / 2 - K(b - a) & K  \leq a \\
     (b^2 - K^2) / 2 - K (b - K) & a < K < b \\
    0 & b \leq K
  \end{cases} = \\
  \label{eq:callprice.uniformdistr.cases_for_K}
       e^{-r T} \cdot \begin{cases}
    (b + a) / 2 - K & K  \leq a \\
     (b - K)^2 / \left[2 (b - a)\right] & a < K < b \\
    0 & b \leq K\, .
    \end{cases}
\end{eqnarray}
The case when $K \geq b$ is not interesting (the price of a call option would be zero) and so assume now that $ K < b$, then the price of the option is
\begin{equation}\label{eq:unifordistr.callprice}
    c(K;\, a,\, b) = \frac{e^{-r T}}{b - a} \cdot \left[ b - \max(K,\, a) \right] \cdot \left[ b/2 + \max(K,\, a) /2 - K \right]\, .
\end{equation}
\par The expected value is equal to ATMF strike\footnote{Otherwise arbitrage opportunity is created.} and is computed as follows:
\begin{equation}\label{eq:ATMF.uniform}
   ATMF = \int_{a}^{b} S_{T} \cdot p(S_T;\, a,\, b) d S_T = \frac{a + b}{2} \, .
\end{equation}
Using put-call parity \eqref{eq:put_call_parity}, the price for the put option is:
\begin{equation}\label{eq:unifordistr.putprice}
    \mbox{put}(K;\, a,\, b) = \frac{e^{-r T}}{b - a} \cdot \left[ b - \max(K,\, a) \right] \cdot \left[ b/2 + \max(K,\, a) /2 - K \right] - e^{-r T} \left( \frac{a + b}{2} - K \right) \, .
\end{equation}

\par
\begin{figure}[h!]
\centering
\includegraphics[scale=0.75]{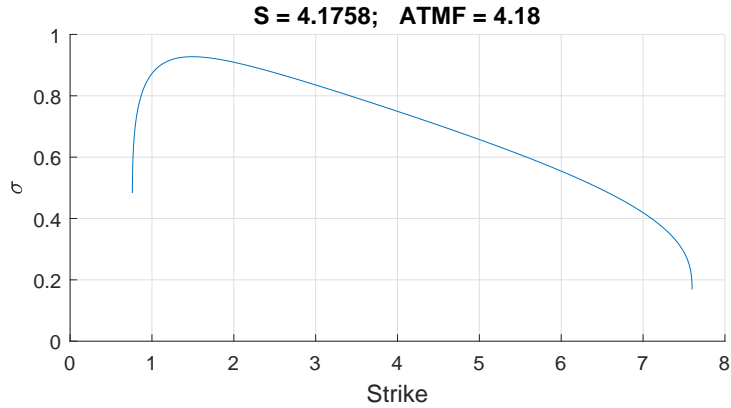} 
\caption{Risk-neutral probability of future asset price is uniform. Volatility smile for vanilla options expiring in 6 month. The profile of the graph is unrealistic for market practice as "real life" implied volatility smiles are convex (with positive 2-nd derivative wrt strike) and not concave as in this graph. The tail of uniform distribution is zero outside its support, this causes such an abrupt concavity of the smile profile.}
\label{fig:smile.uniform}
\end{figure}

 An example of volatility smile corresponding to uniform risk-neutral probability density is depicted in Figure \ref{fig:smile.uniform}.

\subsection{Log uniform distribution}
\par Probability density function for log uniform distribution on an interval $[\ln a,\, \ln b]$, $0 < a < b$, is defined as follows:
\begin{equation}\label{eq:loguniform.distribution}
 p(X;\, a,\, b) =
      \begin{cases}
   \frac{1}{X (\ln b - \ln a)} &  a \leq X \leq  b \\
   0       & \mbox{otherwise}
  \end{cases}
\end{equation}
and corresponds to uniform distribution of $\ln X$.
Using equation \eqref{eq:callprice}, the corresponding call price can be written as:
\begin{eqnarray}\nonumber
    c(K;\, a,\, b) =  \frac{e^{-r T}}{\ln(b / a)} \cdot
    \begin{cases}
    \int_{a}^{b}\left(S_{T} - K \right) / S_{T} d S_T & K  \leq  a \\
     \int_{K}^{b}\left(S_{T} - K \right) /S_{T}  d S_T &  a < K < b \\
    0 & b \leq K
  \end{cases} = \\
  \nonumber
     \frac{e^{-r T}}{\ln(b / a)} \cdot \begin{cases}
    b - a - K\ln(b / a) & K  \leq a \\
     b - K - K \ln(b / K) &  a < K < b \\
    0 & b \leq K
  \end{cases} .
\end{eqnarray}
The case when $K \geq b$ is not interesting (the price of a call option would be zero) and so assume now that $ K < b$, then the price of a call option is
\begin{equation}\label{eq:logunifordistr.callprice}
    c(K;\, a,\, b) = \frac{e^{-r T}}{\ln(b / a)} \cdot \left[ b - \max(K,\, a)  - K \ln(b / \max(K,\, a))\right]\, .
\end{equation}
\par The expected value corresponding to ATMF strike is computed as follows:
\begin{equation}\label{eq:ATMF.loguniform}
   ATMF = \int_{a}^{b} S_{T} \cdot p(S_T;\, a,\, b) d S_T = \frac{b - a}{\ln\left(b/a\right)} \, .
\end{equation}
\par
\begin{figure}[h!]
\centering
\includegraphics[scale=0.75]{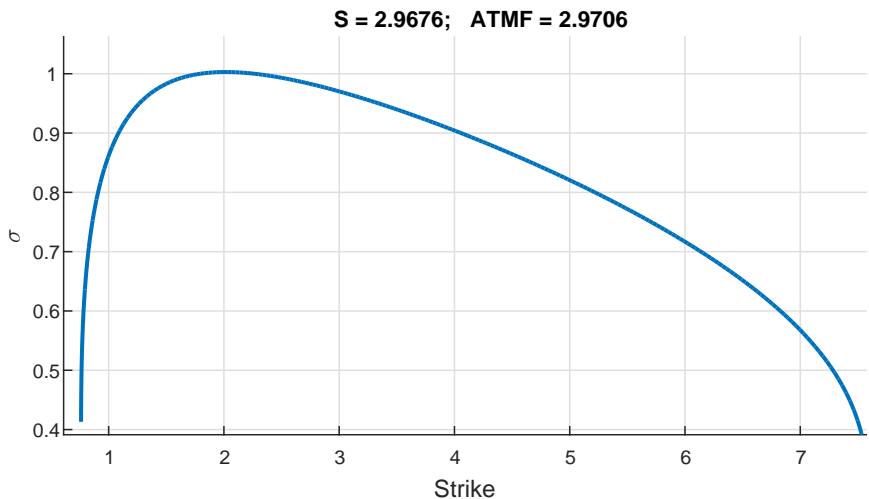} 
\caption{Risk-neutral probability of future asset price is log uniform. Volatility smile for vanilla options expiring in 6 month. The profile of the graph is unrealistic for market practice as "real life" implied volatility smiles are convex (with positive 2-nd derivative wrt strike) and not concave as in this graph. The tail of log-uniform distribution is zero outside its support, this causes such an abrupt concavity of the smile profile.}
\label{fig:smile.log.uniform}
\end{figure}

 An example of volatility smile corresponding to log-uniform risk-neutral probability density is depicted in Figure \ref{fig:smile.log.uniform}.

\subsection{Mixture of log-normal distributions}
\par All distributions considered above but uniform have probability density concentrated mostly near their single maximum. Now consider a different case of a probability distribution whose density may have several points of local maximum.
\par Mixture of log-normal distributions is defined on the same basis as the mixture of normal distributions. Let $\{X_i\}, \; i = 1,\,\ldots,\, n$ be normally distributed independent random variables with expectation $\mu_i$ and standard deviation $\sigma_i$ respectively. Given $n$ non-negative numbers $\{q_i\}, \; i = 1,\,\ldots,\, n$ that sum up to 1, $\sum_{i=1}^n q_i = 1$, the mixture of $n$ log-normal distributions is known to be:
\begin{equation}\label{eq:mixture.log.normals}
  Z = \sum_{i = 1}^{n} q_i e^{X_i}\, .
\end{equation}

\par Denote expectation of each log-normal distribution as follows: $M_i = e^{\mu_i + {\sigma_i}^2/2}$. Direct computations lead to the values of the expectation and variance of $Z$:
\begin{equation}\label{eq:expectation.mixture.log.normals}
  \mathds{E}(Z) \equiv M = \sum_{i=1}^n q_i \cdot M_i = \sum_{i=1}^n q_i \cdot e^{\mu_i + {\sigma_i}^2/2} \, ,
\end{equation}
 Noting that the random variables $\left\{e^{X_i}\right\}$ are also independent,
\begin{eqnarray}
\nonumber
  \mathds{V}ar(Z) & = & \mathds{E}\left[\sum_{i=1}^n q_i  e^{X_i} - M \right]^2 = \mathds{E}\left[\sum_{i=1}^n q_i   \left(e^{X_i} - M_i\right)\right]^2 = \\
  & &
  \label{eq:std.mixture.log.normals}
  \mathds{E}\left[\sum_{i=1}^n {q_i}^2   \left(e^{X_i} - M_i\right)^2 + \sum_{i\neq j} q_i q_j \left(e^{X_i} - M_i\right)  \left(e^{X_j} - M_j\right) \right] = \sum_{i=1}^n {q_i}^2
  \mathds{V}ar\left( e^{X_i} \right) \,.
\end{eqnarray}
\par Now use formula \eqref{eq:callprice} for computing the price of the call option
\begin{eqnarray}
 \nonumber 
   c(K, S, T) & = & e^{-r T} \int_{K}^{\infty}\left(S_{T} - K \right) \cdot p(S_T) d S_T = e^{-r T} \int_{K}^{\infty}\left(S_{T} - K \right) \cdot \sum_{i=1}^n q_i \cdot p(S_T) d S_T = \\
              &   &  \sum_{i=1}^n q_i c_{i}(K, S, T)  \, ,
   \label{eq:callprice.mixture.lognormals}
\end{eqnarray}
where $c_{i}(K, S, T)$ corresponds to call option whose underlying at time $T$ is distributed according to $e^{X_i}$.

\section{Discussion}
\par This work demonstrated examples of how probability distributions can be represented in terms of their corresponding implied volatilities. Of course, only distributions whose integral \eqref{eq:callprice} exists are relevant. I propose the following advantages of representing probability distribution with corresponding volatility smile over probability density function:
\begin{enumerate}
    \item Such representation shows deviation of tails of the probability distribution from log-normal distribution and could visually provide information about fatness of tails versus log-normal distribution.
    \item When empirical data are fitted with  probability density function from prescribed family (e.g. gamma, normal, lognormal) the fit cannot be precise. The deviation could be described by the geometric form of the volatility smile, although such process would require development of a suitable numerical procedure.
    \item It is more natural for humans performing repeatable creative processes that involve ``creative'' measuring of uncertainty, like options' trading.
\end{enumerate}
The above list is, probably, incomplete and further study would contribute to it.
\par Among different assets, interest rates may take negative values. Log-normal distribution is supported only for positive arguments though. In case of possible negative values of an asset, pricing formula \eqref{eq:normal.call.price} with normal implied volatility is used. It could be interesting to see how different probability distributions are represented with normal implied volatility what is left for future analysis.

\appendix
\section{Delta of a call option when future asset price follows gamma distribution}

\par Differentiate equation \eqref{eq:callprice.gamma} with respect to spot $S_0$ in order to get formula for a delta of a call option. First of all find derivative of the gamma probability density function \eqref{eq:gamma.distribution} with respect to parameter $\theta$.
%
\begin{eqnarray}
    \nonumber
  \frac{\partial p(X;\, \hat{\kappa},\, \theta)}{\partial \theta} &=& \frac{X^{\hat{\kappa} - 1} e^{-X/\theta}}{\theta^{\hat{\kappa}} \Gamma(\hat{\kappa})} \cdot \frac{X}{\theta^2} - \hat{\kappa} \cdot \frac{X^{\hat{\kappa} - 1} e^{-X/\theta}}{\theta^{\hat{\kappa} + 1} \Gamma(\hat{\kappa})} =  \\
  \nonumber
   & & \frac{1}{\theta} \cdot \frac{X^{\hat{\kappa}} e^{-X / \theta} \cdot \hat{\kappa}}{\theta \cdot \theta^{\hat{\kappa}} \Gamma(\hat{\kappa}) \cdot \hat{\kappa}} - \frac{\hat{\kappa}}{\theta} \cdot \frac{X^{\hat{\kappa} - 1} e^{-X/\theta}}{\theta^{\hat{\kappa}} \Gamma(\hat{\kappa})} = \\
   \label{eq:partialp_partialtheta}
   & & \frac{\hat{\kappa}}{\theta} \left[p(X;\, \hat{\kappa} + 1,\, \theta) - p(X;\, \hat{\kappa},\, \theta) \right]\; .
\end{eqnarray}
Therefore, noting that cumulative distribution results from integrating the probability density with respect to $X$, $P(K;\, \hat{\kappa},\, \theta) = \int_0^K p(X;\, \hat{\kappa},\, \theta) dX$, differentiation of $P$ with respect to $\theta$ is identical to differentiation of $p$ in formula \eqref{eq:partialp_partialtheta} because the integral has ``nice'' convergence properties.
\par Computations of call's delta will use either of the two assumptions and each assumption will underlie a different value of $\Delta$.
\begin{enumerate}
    \item The parameter $\kappa$ that defines the shape     of the probability density function is constant while only $\theta$ is influenced by the changes in spot $S_{0}$.
    \item Variance of the distribution is constant and both $\theta$ and $\kappa$ are influenced by the changes in spot $S_0$.
\end{enumerate}

\subsection{$\Delta$ under the assumption of constant $\kappa$}

\par The following property is satisfied when $\kappa = \mbox{const}$.
\begin{equation}\label{eq:dtheta.to.ds.when.kappa.constant}
    \left.\frac{d\theta}{d S_0}\right|_{\kappa = \mbox{const}} = \left.\frac{d \left(S_{0} e^{(r_T - q_T) T} / \kappa\right)}{d S_0}\right|_{\kappa = \mbox{const}} = \frac{e^{(r_T - q_T) T}}{\kappa} \, .
\end{equation}

Now use formulae \eqref{eq:partialp_partialtheta} and \eqref{eq:dtheta.to.ds.when.kappa.constant} to differentiate cumulative gamma probability with respect to spot $S_0$:
\begin{eqnarray}
    \nonumber
    \frac{d}{d S_0} P(K;\, \hat{\kappa},\, \theta(S_0,\, \hat{\kappa})) & = &  \frac{\partial}{\partial \theta} P(K;\, \hat{\kappa},\, \theta) \cdot \frac{d}{d S_0} \theta = \\
    \nonumber
    & & \frac{\hat{\kappa}}{\theta} \left[P(K;\, \hat{\kappa} + 1,\, \theta) - P(K;\, \hat{\kappa},\, \theta) \right] \cdot e^{\left(r_{T} - q_T\right) T}  / \kappa = \\
    \label{eq:dPgamma_dS0}
     & & \frac{\hat{\kappa} \cdot e^{\left(r_{T} - q_T\right) T} }{\kappa \cdot \theta} \left[P(K;\, \hat{\kappa} + 1,\, \theta) - P(K;\, \hat{\kappa},\, \theta) \right] \; .
\end{eqnarray}
Here $\hat{\kappa}$ is an arbitrary parameter of the probability function while $\kappa = \mbox{ATMF} / \theta$ from \eqref{eq:ATMF.gamma} corresponds to the risk-neutral probability density of the future asset price.
Now use formula \eqref{eq:dPgamma_dS0} to differentiate the call price \eqref{eq:BScall.S0.insteadof.theta} with respect to $S_0$.
\begin{eqnarray}
    \nonumber
    \frac{d c}{d S_0} & = & e^{-q_T T} \cdot \left[1 - P(K; \kappa + 1, \theta) \right] + \\
    \nonumber
    & & e^{-r_T T} \frac{e^{(r_T - q_T) T}}{\kappa \cdot \theta} \left\{ - S_{0}\cdot (\kappa + 1) \cdot e^{\left(r_{T} - q_T\right) T}  \cdot \left[P(K;\, \kappa + 2,\, \theta) - P(K;\, \kappa + 1,\, \theta) \right] \right. + \\
    \nonumber
    & &  \left. K \cdot \kappa \cdot  \left[P(K;\, \kappa + 1,\, \theta) - P(K;\, \kappa,\, \theta) \right] \right\} = \\
    \nonumber
    & &  e^{-q_T T} \cdot \left\{ 1 - P(K; \kappa + 1, \theta)  + \right. \\
     \nonumber
    & & \left . \frac{K}{\theta} \cdot \left[P(K;\, \kappa + 1,\, \theta)- P(K;\, \kappa,\, \theta) \right]  - \frac{\mbox{ATMF} \cdot (\kappa + 1)}{\kappa \cdot \theta} \cdot \left[P(K;\, \kappa + 2,\, \theta) - P(K;\, \kappa + 1,\, \theta) \right] \right\} = \\
     & &
     \nonumber
      e^{-q_T T} \cdot \left\{ 1 - P(K; \kappa + 1, \theta)  +       \frac{K}{\theta} \cdot \left[P(K;\, \kappa + 1,\, \theta) - P(K;\, \kappa,\, \theta) \right]  - \right. \\
      \nonumber
      & & \left. (\kappa + 1) \cdot \left[P(K;\, \kappa + 2,\, \theta) - P(K;\, \kappa + 1,\, \theta) \right] \right\} = \\
      \label{eq:dc.dS0}
      & &  e^{-q_T T} \cdot \left\{ 1 -  K \cdot P(K;\, \kappa,\, \theta) / \theta - (\kappa + 1) P(K;\, \kappa + 2,\, \theta)  +  P(K; \kappa + 1, \theta) \left[ \frac{K}{\theta} + \kappa  \right] \right\}\, ,
\end{eqnarray}
formula \eqref{eq:ATMF.gamma} was used in the equality one before the last.

\subsection{$\Delta$ under the assumption of constant variance}
\par Use formulae (\ref{eq:thetagamma.of.variance.mean}, \ref{eq:kappa.of.variancemean}) to compute derivatives $d\kappa/d S_0$ and $d\theta/d S_0$ under the assumption of constant variance:
\begin{eqnarray}
    \label{eq:dkappa.to.dS0.given.constantvariance}
    \left.\frac{d \kappa}{d S_0}\right|_{\sigma^2 = \kappa\theta^2 = \mbox{const}} \equiv \kappa' &=& 2\frac{{S_0} e^{2(r_T - q_T) T}}{\sigma^2} = - \frac{2 e^{(r_T - q_T) T}}{S_0 \theta'}\, , \\
    \label{eq:dtheta.to.dS0.given.constantvariance}
  \left.\frac{d \theta}{d S_0}\right|_{\sigma^2 = \kappa\theta^2 = \mbox{const}} \equiv \theta' &=& -\frac{\sigma^2}{{S_0}^2 e^{(r_T - q_T) T}}  = - 2 S_0 e^{(r_T - q_T) T} / \kappa' = -2 \mbox{ATMF} / \kappa'   \, .
\end{eqnarray}

Now find the partial derivative
\begin{equation}\label{eq:partialp.to.partialkappa}
    \frac{\partial p(X; \hat{\kappa}, \hat{\theta})}{\partial\hat{\kappa}} = \frac{X^{\hat{\kappa} - 1} e^{-\frac{X}{\hat{\theta}}}}{\hat{\theta}^{\hat{\kappa}} \Gamma(\hat{\kappa})} \cdot \left[(\hat{\kappa} - 1) \ln X - \hat{\kappa} \ln\hat{\theta} - \psi(\hat{\kappa}) \right] = \left[(\hat{\kappa} - 1) \ln X - \hat{\kappa} \ln\hat{\theta} - \psi(\hat{\kappa}) \right] \cdot p(X; \hat{\kappa}, \hat{\theta})\, ,
\end{equation}
$\psi(\kappa) = \frac{d\Gamma(\kappa)/d\kappa}{\Gamma(\kappa)}$ is the digamma function.
There is no functional form that relates the integral of the right hand side of \eqref{eq:partialp.to.partialkappa} over $X$ to cumulative probability function $P$ due to the factor $\ln X$, in contrary to the partial derivative of $p$ with respect to $\theta$ in \eqref{eq:partialp_partialtheta}. For this reason functional expression of $\Delta$ for the case of constant variance is not derived. Numeric computation of $\Delta$ can be based on numerical differentiation of the expression

\begin{align}
\nonumber
  \left.c(K, r, T)\right|_{\sigma^2 = \mbox{const}} & =    e^{-r_T T} \left\{ S_{0}\cdot e^{\left(r_{T} - q_T\right) T}  \cdot \left[1 - P(K; \frac{{S_0}^2 e^{2 (r_T - q_T) T}}{\sigma^2} + 1, \frac{\sigma^2}{S_0 e^{(r_T - q_T) T}}) \right] - \right. \\
  \label{eq:BScall.variance.insteadof.kappatheta}
   &  \left. K \cdot \left[1 - P(K; \frac{{S_0}^2 e^{2 (r_T - q_T) T}}{\sigma^2}, \frac{\sigma^2}{S_0 e^{(r_T - q_T) T}}) \right] \right\}
\end{align}
with respect to $S_0$, where $\kappa$ and $\theta$ from \eqref{eq:BScall.ATMF.insteadof.kappatheta} were replaced based on formulae \eqref{eq:thetagamma.of.variance.mean}, \eqref{eq:kappa.of.variancemean}.

\section{Delta of a call option when future asset price follows translated Student $t$ distribution}

\par Differentiate equation \eqref{eq:call.student.final} with respect to spot $S_0$ in order to get formula for a delta of a call option. First of all note the expression for derivative of the regularized incomplete beta function $I_{x}$ from \eqref{eq:incomplete.beta.function} with respect to $x$:
\begin{equation}\label{eq:dIx.dx}
   \frac{d I_x(a,\, b)}{d x} = \frac{x^{a - 1} (1 - x)^{b - 1}}{B(a,\, b)} \; .
\end{equation}
So, for $y(x)$ from \eqref{y_of_x}
\begin{multline}\label{eq:dIyx.dmu}
  \frac{d I_{y(K)}(a,\, b)}{d \mu} = \frac{d y(K)}{d\mu} \cdot \frac{d I_y(a,\, b)}{d y} = -2 \frac{(\mu - K) y(K)^2}{\nu} \cdot \frac{y(K)^{a - 1} \left[1 - y(K)\right]^{b - 1}}{B(a,\, b)} = \\
  -2 \frac{(\mu - K) \nu^{2 + a - 1} (\mu - K)^{2 (b - 1)}}{\nu \left[(\mu - K)^2 + \nu \right]^{2 + a - 1 + b - 1} B(a,\, b)}  =
  -2 \frac{ (\mu - K) \cdot |\mu - K|^{2 (b - 1)} \nu^a}{\left[(\mu - K)^2 + \nu \right]^{a + b} B(a,\, b)} \; .
\end{multline}
Correspondingly, from \eqref{eq:dIyx.dmu}
\begin{equation}\label{eq:dIyx.dmu.nuover2.onehalf}
   \frac{d}{d\mu} I_{y(K)}\left(\frac{\nu}{2},\, \frac{1}{2}\right) = -2 \frac{ (\mu - K) \cdot \nu^{\nu/2}}{|\mu - K| \cdot \left[(\mu - K)^2 + \nu \right]^{(\nu + 1)/2} B\left(\frac{\nu}{2},\, \frac{1}{2}\right)} \; .
\end{equation}
So, using formula \eqref{eq:dIyx.dmu.nuover2.onehalf}, the formula for delta of the call option \eqref{eq:call.student.final} is:
\begin{multline}\label{eq:deltacall.student}
  \Delta = \frac{d c}{d S_0} = \frac{d c}{d \mu} \cdot \frac{\partial \left[S_0 e^{(r - q) T}\right]}{{\partial S_0}}  =        - e^{-q T}  \frac{\Gamma\left( \frac{\nu + 1}{2} \right)}{\sqrt{\nu \pi}\, \Gamma \left( \frac{\nu}{2} \right)} \cdot \left(\mu - K \right) \cdot \left[ 1 + \frac{(\mu - K)^2}{\nu} \right]^{-\frac{1 + \nu}{2}} + \\ \frac{e^{-q T}}{2} \cdot
     \begin{cases}
        I_{y(K)} \left( \frac{\nu}{2},\, \frac{1}{2}\right) - 2 \frac{\nu^{\nu/2} \cdot |\mu - K|}{\left[(\mu - K)^2 + \nu \right]^{(\nu + 1)/2} B\left(\frac{\nu}{2},\, \frac{1}{2}\right)}    & K  \geq  \mu \\
        & \\
     2 -\left[I_{y(K)} \left( \frac{\nu}{2},\, \frac{1}{2}\right) - 2 \frac{\nu^{\nu/2} \cdot |\mu - K|}{\left[(\mu - K)^2 + \nu \right]^{(\nu + 1)/2} B\left(\frac{\nu}{2},\, \frac{1}{2}\right)}\right]   &  K < \mu
  \end{cases} \; .
\end{multline}

\section{Delta of a call option when future asset price follows uniform distribution}
\par Use \eqref{eq:ATMF.uniform} to obtain $S_0 = e^{-(r - q) T} \mbox{ATMF} = e^{-(r - q) T} \frac{a - b}{2}$. So,
\begin{equation}\label{eq:b.versus.S_0.uniformdistr}
  b = 2 S_0 e^{(r - q) T} - a\, .
\end{equation}
Assume at the moment that $a < K < b$ and substitute \eqref{eq:b.versus.S_0.uniformdistr} into \eqref{eq:callprice.uniformdistr.cases_for_K} to get
\begin{equation}\label{eq:callprice.versus.S_0.uniformdistr}
  c(K;\, S_0,\, a) = e^{-r T} \frac{\left(2 S_0 e^{(r - q) T} - a - K \right)^2}{4 (S_0 e^{(r - q) T} - a)} \, .
\end{equation}
Now compute $\Delta$:
\begin{eqnarray}
 \nonumber 
  \Delta &=& \frac{d c}{d S_0} = e^{-r t} \frac{2 \cdot 2 e^{(r - q) T} \left(2 S_0 e^{(r - q) T} - a - K \right) \cdot 4 (S_0 e^{(r - q) T} - a) -  \left(2 S_0 e^{(r - q) T} - a - K \right)^2   \cdot 4 \cdot e^{(r - q) T}}{[4 (S_0 e^{(r - q) T} - a)]^2} =  \\
  \nonumber
   & & 4 e^{-r t} e^{(r - q) T} (2 S_0 e^{(r - q) T} - a - K) \frac{4 S_0 e^{(r - q) T} - 4 a - 2 S_0 e^{(r - q) T} + a + K}{16 (S_0 e^{(r - q) T} - a)^2} =  \\
   \label{eq:delta.call.uniformdistr.intermediate}
   & & 4 e^{- q T} \frac{(2 S_0 e^{(r - q) T} - a - K) \cdot( 2 S_0 e^{(r - q) T} - 3 a + K)}{16 (S_0 e^{(r - q) T} - a)^2} \, .
\end{eqnarray}
Equations \eqref{eq:delta.call.uniformdistr.intermediate} and \eqref{eq:b.versus.S_0.uniformdistr} imply:
\begin{eqnarray}\nonumber 
  \Delta &=& 4 e^{- q T} \frac{(b - K) \cdot( b- 2 a + K)}{4 (b - a)^2} = e^{- q T} \frac{(b - K) \cdot( b- 2 a + K)}{(b - a)^2} = \\
  \label{eq:delta.call.uniformdistr}
   &=& e^{- q T} \frac{(b - a)^2 - (K - a)^2}{(b - a)^2} = e^{- q T} \left(1 - \frac{(K - a)^2}{(b - a)^2} \right) \, .
\end{eqnarray}

Now assume that $K \leq a$ and substitute \eqref{eq:b.versus.S_0.uniformdistr} into \eqref{eq:callprice.uniformdistr.cases_for_K} to get
\begin{equation}\label{eq:callprice.versus.S_0.uniformdistr.case_intermediate_K}
  c(K;\, S_0,\, a) = e^{-r T} \left( S_0 e^{(r - q) T} - K \right)
\end{equation}
and
\begin{equation}\nonumber
  \Delta = \frac{d c}{d S_0} = e^{-q T} \, .
\end{equation}

So finally, the formula for $\Delta$ of the call option when asset price is uniformly distributed is as follows.

\begin{eqnarray}
    \nonumber
    \Delta = e^{-q T}   \cdot
    \begin{cases}
        1 & K  \leq a \\
        1 - \frac{(K - a)^2}{(b - a)^2} & a < K < b \\
        0 & b \leq K
    \end{cases} = \\
    \label{eq:delta.call.uniformdistr.allcases}
    \begin{cases}
        e^{-q T}\left[1 - \frac{(\max(K,\, a) - a)^2}{(b - a)^2} \right]  & K  < b \\
        0 & b \leq K \, .
    \end{cases}
\end{eqnarray}
Correspondingly, for $a < K < b$,
\begin{equation}\label{eq:K.from.delta.uniformdistr}
  K = a + (b - a) \sqrt{1 - \Delta \cdot  e^{q T}}   \, .
\end{equation}

\bibliographystyle{plain}
\bibliography{xbib_all}
\end{document}